\documentclass[reprint,
 amsmath,amssymb,
 aps,
floatfix,
showkeys
]{revtex4-1}

\usepackage[USenglish]{babel}
\usepackage{slashed}
\usepackage{physics}
\usepackage{graphicx}
\usepackage{epstopdf}
\usepackage{graphics}
\usepackage{dcolumn}
\usepackage{bm}
\usepackage[usenames]{color}
\usepackage[svgnames]{xcolor}
\usepackage{comment}
\usepackage[colorlinks,citecolor=DarkGreen,linkcolor=DarkRed,urlcolor=DarkBlue]{hyperref}
\usepackage{float}

\newcommand{\be}{\begin{equation}}
\newcommand{\ee}{\end{equation}}
\newcommand{\bea}{\begin{eqnarray}}
\newcommand{\eea}{\end{eqnarray}}
\newcommand{\beas}{\begin{eqnarray*}}
\newcommand{\eeas}{\end{eqnarray*}}

\begin{document}

\title{Magnetic field dependence of the neutral pion longitudinal screening mass in the linear sigma model with quarks}


\author{Alejandro Ayala$^{1,2}$, Ricardo L. S. Farias$^2$, L. A. Hern\'andez$^{3}$, Ana Julia Mizher$^{4,5,6}$, Javier Rend\'on$^1$, Cristian Villavicencio$^{6}$, R. Zamora$^{7,8}$}
\affiliation{%
$^1$Instituto de Ciencias Nucleares, Universidad Nacional Aut\'onoma de M\'exico, Apartado Postal 70-543, CdMx 04510, Mexico.\\
$^2$Departamento de F\'isica, Universidade Federal de Santa Maria, Santa Maria, RS 97105-900, Brazil.\\
$^3$Departamento de F\'isica, Universidad Aut\'onoma Metropolitana-Iztapalapa, Avenida San Rafael Atlixco 186, CdMx 09340, Mexico.\\
$^4$ Instituto de F\' isica Te\'orica, Universidade Estadual Paulista, Rua Dr. Bento Teobaldo Ferraz, 271 - Bloco II, 01140-070 S\~ao Paulo, SP, Brazil.\\
$^5$Laboratório de Física Teórica e Computacional, Universidade Cidade de São Paulo, 01506-000, São Paulo, Brazil.\\
$^6$Centro de Ciencias Exactas and $^8$Departamento de Ciencias B\'asicas, Facultad de Ciencias, Universidad del B\'io-B\'io, Casilla 447, Chill\'an, Chile.\\
$^7$Instituto de Ciencias B\'asicas, Universidad Diego Portales, Casilla 298-V, Santiago, Chile.\\
$^8$Facultad de Medicina Veterinaria, Universidad San Sebastián, Santiago, Chile.}


\begin{abstract}

We use the linear sigma model with quarks to study the magnetic-field-induced modifications on the longitudinal screening mass for the neutral pion at one-loop level. The effects of the magnetic field are introduced into the self-energy, which contains the contributions from all the model particles. We find that, to obtain a reasonable description for the behavior with the field strength, we need to account for the magnetic field dependence of the particle masses. We also find that the couplings need to decrease fast enough with the field strength to then reach constant and smaller values as compared to their vacuum ones. The results illustrate the need to treat the magnetic corrections to the particle masses and couplings in a self-consistent manner, accounting for the backreaction of the field effects for the magnetic field dependence of the rest of the particle species and couplings in the model.

\end{abstract}


\maketitle

\section{Introduction}\label{sec1}
In recent years, it has become clear that electromagnetic fields provide a powerful probe to explore the properties of the QCD vacuum. When the energy associated with the field strength is larger than $\Lambda_{QCD}$, the field can prove the hadron structure and help reveal the dynamics associated with confinement and chiral symmetry breaking. For example, at zero temperature, magnetic fields catalyze the breaking of chiral symmetry, producing a stronger light quark-antiquark condensate~\cite{catalysis}. However, for nonvanishing temperature, magnetic fields inhibit the condensate formation and reduce the critical temperature for chiral symmetry restoration, giving rise to inverse magnetic catalysis (IMC)~\cite{LQCD,Bruckmann,Farias,Ferreira,Ayala0,Ayala2,Ayala3,Avancini,Ayala4,vertex1,vertex2,qcdcoupling,Mueller,imcreview}. This property has motivated an intense activity aimed to search for the influence of magnetic fields on hadron dynamics~\cite{bali01,simonov03,aguirre02,tetsuya,dudal04,kevin,gubler,noronha01,morita,morita02,sarkar03,band,Ayalachi,nosso1,zhuang,scoccola01,huang01,scoccola02,scoccola03,luch,farias01,mao01,sarkar,sarkar02,zhang01,huan02,simonov01,fraga01,aguirre,taya,shinya,andersen01,kojo,loewe1,loewe2,hernandez1,simonov02,ghoshrho,AMMmesons,TBspectralprop,ghoshEPJA,Avila,Avila:2020ved,dudal,dudal02,andrei02,he,nucleon,barionslattice,sr,FESR3,gA_B}. Since the dynamics of chiral symmetry breaking is dominated by pions, the lightest of all quark-antiquark bound states, it then becomes important to explore how the pion mass is affected by the presence of magnetic fields. 

Recall that, for a Lorentz-invariant system, the mass corresponds to the rest energy of a given particle, which can then be obtained from the pole of the propagator when the particle three-momentum $\vec{q}$ is taken to zero. This is dubbed the ``{\it pole mass}''.  Notice that if, instead, the zeroth component of the particle four-momentum $q_0$ is taken first to zero, we obtain the ``{\it screening mass}''. The screening mass squared can be identified as the negative of the particle magnitude of its three-momentum squared. In a system where Lorentz symmetry is unbroken, the pole and screening masses coincide. However, when Lorentz symmetry is broken, as is, for example, the case of a system at finite temperature, the above described limiting procedures do not yield the same values. Explicitly, if $f(q_0,|\vec{q}|;T)$ represents the thermal medium response function that contributes to the particle dispersion relation, the limits
$f(q_0,0;T)$ and $f(0,|\vec{q}|;T)$ do not commute. Pole and screening masses do not coincide. The name screening mass stems from the analysis in linear response theory when studying the influence of static external fields on a thermal medium. Because of the static nature of the external field, its screening within the medium is controlled by the system's response function in the limit $f(0,|\vec{q}|;T)$. The inverse of the screening mass corresponds to the screening or Debye length.

When the system is immersed in a magnetic field, the breaking of Lorentz symmetry happens in the spatial directions, giving rise to distinct dispersion properties for particles moving in the transverse or the longitudinal directions with respect to the field orientation. Thus, in addition to studying the magnetic-field-induced modifications of the pole mass, one can also study the corresponding longitudinal and transverse screening masses. At $T=0$ the longitudinal screening mass is equal to the pole mass. This degeneracy is lifted when $T\neq 0$. Although most of the studies have concentrated in the magnetic properties of the pion pole mass~\cite{Daspole,Avancinipole1,AHHFZ,Ayala1,nosso03,FESR2,Hidaka,Luschevskaya,Endrodi,Ding,Das,Li,Scoccola}, more recently, an interesting relation between the magnetic behavior of screening masses and condensates, and thus between IMC and screening masses, has been obtained in Ref.~\cite{Ding}. Motivated by this finding, the authors of Ref.~\cite{Ding-2022} used a lattice QCD (LQCD) setup to assess the importance of the ``{\it sea}'' versus the ``{\it valence}'' quarks' contribution for the temperature and magnetic dependence of the pion screening mass. For the lowest temperature studied, the screening mass seems to behave as a monotonically decreasing function of the field strength, up to $|eB|\sim 2.5$ GeV$^2$. Unfortunately, no attempt to distinguish between longitudinal and transverse screening masses was made. The transverse and longitudinal pion masses at finite temperature and magnetic field strength were also studied in Ref.~\cite{Sheng} using a two-flavor Nambu--Jona-Lasinio (NJL) model in the random phase approximation. The authors focused on addressing possible mishaps of previous calculations~\cite{Fayazbakhsh1,Fayazbakhsh2}. Their results indicate opposite behaviors for the transverse and longitudinal screening masses as functions of the magnetic field strength for $T=0$; whereas the former decreases, the latter slightly increases.

Since it is important to check that in a magnetic background the pole and the longitudinal screening mass are equal when calculated within the framework of a given effective model at $T=0$, in this work we use the linear sigma model with quarks (LSMq) to study the pion longitudinal screening mass as a function of a magnetic field of arbitrary strength for vanishing temperature. We argue that to extract a reliable behavior of this mass as a function of the field strength, the magnetic field dependence of the couplings, as well as of the quark, pion, and $\sigma$ pole masses, need to be accounted for and that, in this sense, the complete solution of the particle mass dependence on the magnetic field needs to be treated self-consistently within a given model. We find that a rapid decrease of the model couplings with the field strength is needed for the longitudinal screening mass to follow the LQCD profile as a function of the magnetic field. This procedure is consistent with previous calculations of the magnetic field dependence of the pion pole mass within the same model, where it was also found that a rapid reduction of the couplings with the field strength is needed to describe the magnetic field behavior of the pion pole mass~\cite{AHHFZ,Ayala1}. Since the LSMq provides a general framework to study quark-meson systems under the influence of magnetic fields, the setup can also be extended to address the properties of the directional sound velocities, a subject that is studied in Ref.~\cite{Ding-2022}, or the connection between screening masses and inverse magnetic catalysis, a subject that is emphasized in Ref.~\cite{Sheng}. These studies require, as a previous step, the implementation of an adequate formulation of the magnetic field effects at zero temperature, the subject that we explore in the present work. The work is organized as follows: In Sec.~\ref{sec2}, we introduce the linear sigma model with quarks. In  Sec.~\ref{sec3}, we make a quick survey of the way magnetic field effects are introduced into the propagators of charged bosons and fermions, which we hereby describe in terms of the Schwinger proper time formalism. In  Sec.~\ref{sec4}, we compute the Feynman diagrams that contribute to the neutral pion self-energy. In Sec.~\ref{sec5}, we compute the magnetic corrections to the neutral pion screening masses, showing that the behavior strongly depends on the magnetic field dependence of masses and couplings. 
We finally summarize and conclude in Sec~\ref{sec6}. We reserve for the Appendix the explicit calculation details of the one-loop magnetic filed corrections to the neutral pion self-energy.

\section{Linear Sigma Model with quarks} \label{sec2}
The LSMq is an effective model that describes the low-energy regime of QCD, incorporating the spontaneous breaking of chiral symmetry. The Lagrangian for the LSMq can be written as 
\begin{eqnarray}
\mathcal{L}&=&\frac{1}{2}(\partial_{\mu}\sigma)^{2}+\frac{1}{2}(\partial_{\mu}\vec{\pi})^{2}+\frac{a^{2}}{2}(\sigma^{2}+\vec{\pi}^{2})-\frac{\lambda}{4}(\sigma^{2}
+\vec{\pi}^{2})^{2}\nonumber\\
&+&i\bar{\psi}\gamma^{\mu}\partial_{\mu}\psi-ig\gamma^{5}\bar{\psi} \vec{\tau} \cdot \vec{\pi}\psi-g\bar{\psi}\psi\sigma.
\label{lagrangian}
\end{eqnarray}
Pions are described by an isospin triplet,  $\vec{\pi}=(\pi_1,\pi_2,\pi_3)$. Two species of quarks are represented by an $SU(2)$ isospin doublet $\psi$. The $\sigma$ scalar is included by means of an isospin singlet. Also, $\lambda$ is the boson self-coupling and $g$ is the fermion-boson coupling. $a^2>0$ is the mass parameter.

To allow for spontaneous symmetry breaking, we let the $\sigma$ field develop a vacuum expectation value $v$,
\begin{eqnarray}
    \mathcal{L}&=&\frac{1}{2}\partial_{\mu}\sigma \partial^{\mu}\sigma+\frac{1}{2}\partial_{\mu}\pi_{0}\partial^{\mu}\pi_{0}+\partial_{\mu}\pi_{-}\partial^{\mu}\pi_{+}\nonumber\\
    &-&\frac{1}{2}m_{\sigma}^{2}\sigma^{2}-\frac{1}{2}m_{0}^{2}\pi_{0}^{2}-m_{0}^{2}\pi_{-}\pi_{+}+i\bar{\psi}\slashed{\partial}\psi\nonumber\\
    &-&m_{f}\bar{\psi}\psi+\frac{a^2}{2}v^2-\frac{\lambda}{4}v^4+\mathcal{L}_{int},
    \label{linearsigmamodelSSB}
\end{eqnarray}
where the charged pion fields can be expressed as
\begin{equation}
 \pi_\pm=\frac{1}{\sqrt{2}}(\pi_1\pm i\pi_2),
\end{equation}
and the interaction Lagrangian is defined as
\begin{equation}
\begin{split}
    \mathcal{L}_{int}&=-\frac{\lambda}{4}\sigma^{4}-\lambda v\sigma^{3}-\lambda v^{3}\sigma-\lambda\sigma^{2}\pi_{-}\pi_{+} -2\lambda v \sigma\pi_{-}\pi_{+}\\
    &-\frac{\lambda}{2}\sigma^{2}\pi_{0}^{2}-\lambda v\sigma \pi_{0}^{2}-\lambda \pi_{-}^{2}\pi_{+}^{2}-\lambda\pi_{-}\pi_{+}\pi_{0}^{2}-\frac{\lambda}{4}\pi_{0}^{4}\\ 
    &+a^{2}v\sigma -g\bar{\psi}\psi\sigma-ig\gamma^{5}\bar{\psi}\left(\tau_{+}\pi_{+}+\tau_{-}\pi_{-}+\tau_{3}\pi_{0}\right)\psi.
    \label{interactinglagrangian}
\end{split}    
\end{equation}
In order to include a finite vacuum pion mass $m_{0}$, one adds an explicit symmetry breaking term in the Lagrangian of Eq.~\eqref{linearsigmamodelSSB} such that
\begin{equation}
    \mathcal{L}\rightarrow \mathcal{L'}=\mathcal{L}+hv.
    \label{explicittermLagrangian}
\end{equation}
As can be seen from Eqs.~(\ref{linearsigmamodelSSB}) and~(\ref{interactinglagrangian}), there are new terms that depend on $v$ and all fields develop dynamical masses, 
\begin{align}
     m_{\sigma}^{2}&=3\lambda v^2-a^2, \nonumber \\
     m_{0}^{2}&=\lambda v^2-a^2, \nonumber \\ 
     m_{f}&=gv.
\label{masses}
\end{align}

Using Eqs.~(\ref{linearsigmamodelSSB}) and~(\ref{explicittermLagrangian}), the tree-level potential is given by
\begin{equation}
 V^{\text{tree}}(v)=-\frac{a^2}{2}v^2+\frac{\lambda}{4}v^4-hv.
 \label{treelevel}
\end{equation}
This potential develops a minimum, called the vacuum expectation value of the $\sigma$ field, namely, 
\begin{equation}
    v_{0}=\sqrt{\frac{a^2+m_{0}^2}{\lambda}}.
\label{vev}
\end{equation}
\begin{figure}[t]
\centering
    \includegraphics[scale=0.25]{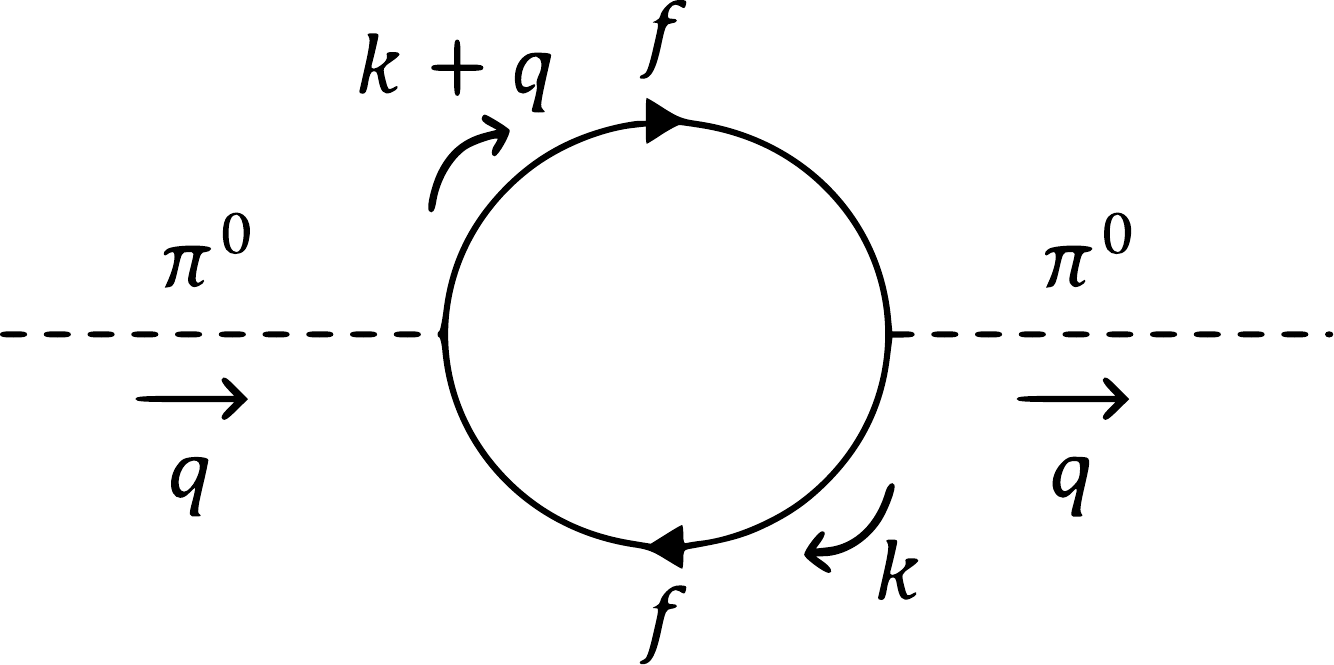}
    \caption{Feynman diagram corresponding to the one-loop contribution from the fermion anti-fermion loop to the neutral pion self-energy in the LSMq.}
    \label{self-energyquarks}
\end{figure}
Therefore, the masses evaluated at $v_0$ are
\begin{align}
 m_f(v_0)&=g\sqrt{\frac{a^2+m_{0}^2}{\lambda}}, \nonumber \\
 m_\sigma^2(v_0)&=2a^2+3m_{0}^2, \nonumber \\
 m_{0}^2(v_0)&=m_{0}^2.
 \label{masses2}
\end{align}
Finally, an external magnetic field, uniform in space and constant in time, can be included in the model introducing a covariant derivative in the Lagrangian density, Eq.~(\ref{linearsigmamodelSSB}), namely, 
\begin{equation}
 \partial_\mu\to D_\mu=\partial_\mu+ieA_\mu,
\end{equation}
where $A^\mu$ is the vector potential corresponding to an external magnetic field directed along the  $\hat{z}$ axis coupled to a particle with charge $e$. In the symmetric gauge, this is given by
\begin{equation} \label{vectorpotencial}
 A^\mu(x)=\frac{1}{2}x_{\nu}F^{\nu\mu},
\end{equation}
and couples only to the charged pions  and to the quarks. 

Notice that, in order to consider the propagation of charged particles, one can resort to introducing Schwinger propagators, which can be expressed either in terms of their proper time representation or as a  sum over Landau levels. For completeness of the presentation, we now proceed to briefly discuss the properties of these propagators.
\begin{figure}[t]
\centering
\includegraphics[scale=1]{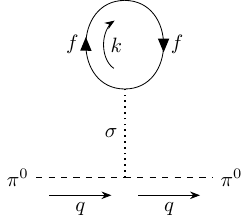}
    \caption{Feynman diagram corresponding to the tadpole contribution from the fermion loop and a sigma to the neutral pion self-energy in the LSMq.}
    \label{tadpole1}
\end{figure}

\section{\label{sec3} Magnetic field-dependent boson and fermion propagators}

To consider the propagation of charged particles within a magnetized background, we use Schwinger's proper time representation. The fermion propagator can be written as~\cite{schwinger}
\begin{equation}
    S_f(x,x')=e^{i\Phi_f(x,x')}S_f(x-x'),
    \label{fermionpropagatorincoordinatespace}
\end{equation}
where $\Phi_f(x,x')$ is the Schwinger phase given by
\begin{eqnarray}
\Phi_f(x,x')=q_f\int_x^{x'}d\xi_\mu \left[
A^\mu(\xi) + \frac{1}{2}F^{\mu\nu}(\xi-x')_\nu
\right],
\label{phase}
\end{eqnarray}
and $q_f$ is the charge of a quark with flavor $f$.
$\Phi_f(x,x')$ corresponds to the translationally noninvariant and gauge-dependent part of the propagator. On the other hand, $S_f(x-x')$ is translationally and gauge invariant and can be expressed in terms of its Fourier transform as
\begin{equation}
    S_f(x-x')=\int \frac{d^{4}p}{(2\pi )^{4}}S_f(p)e^{-ip\cdot(x-x')}, \label{Fouriertransformfermionpropagator}
\end{equation}
where
\begin{eqnarray}
    iS_f(p)&=&\int_0^\infty \frac{ds}{\cos(|q_fB|s)}e^{is\left(p_\parallel^2-p_\perp^2\frac{\tan(|q_fB|s)}{|q_fB|s}-m_f^2+i\epsilon\right)}\nonumber\\
&\times&\left[
\Big(
\cos(|q_fB|s) + \gamma_1\gamma_2\sin(|q_fB|s)\text{sign}(q_fB)
\Big)\right.\nonumber\\
&\times&\left.\left(m_f +\slashed{p}_\parallel\right) - \frac{\slashed{p}_\perp}{\cos(|q_fB|s)}
\right],\label{fermionpropagatormomentumspace}
\end{eqnarray}
with $m_f$ as the quark mass.
In a similar fashion, for a charged scalar field we have  
\begin{eqnarray}
D(x,x')&=&e^{i\Phi_b(x,x')}D(x-x'),\nonumber\\
D(x-x')&=&\int \frac{d^{4}p}{(2\pi)^{4}}D(p)e^{-ip\cdot(x-x')},
\label{scalarprop}
\end{eqnarray}
with
\begin{eqnarray}
iD(p)&=&\int_0^\infty \frac{ds}{\cos(|q_bB|s)}e^{is\left(p_\parallel^2-p_\perp^2\frac{\tan(|q_bB|s)}{|q_bB|s}-m_b^2+i\epsilon \right)},\nonumber\\
\label{bosonpropagatormomentumspace}
\end{eqnarray}
where $m_b$ and $q_b$ are the boson mass and charge, respectively. The $\epsilon$ appearing in Eqs.~(\ref{fermionpropagatormomentumspace}) and~(\ref{bosonpropagatormomentumspace}) is the  infinitesimal positive parameter that enforces Feynman boundary conditions and thus causality. Notice that, in the $B\to0$ limit, one recovers the usual Feynman fermion and scalar propagators.

We now use these ingredients to compute the elements necessary to obtain the magnetic modification of the neutral pion mass.
\begin{figure}[t]
    \includegraphics[scale=0.25]{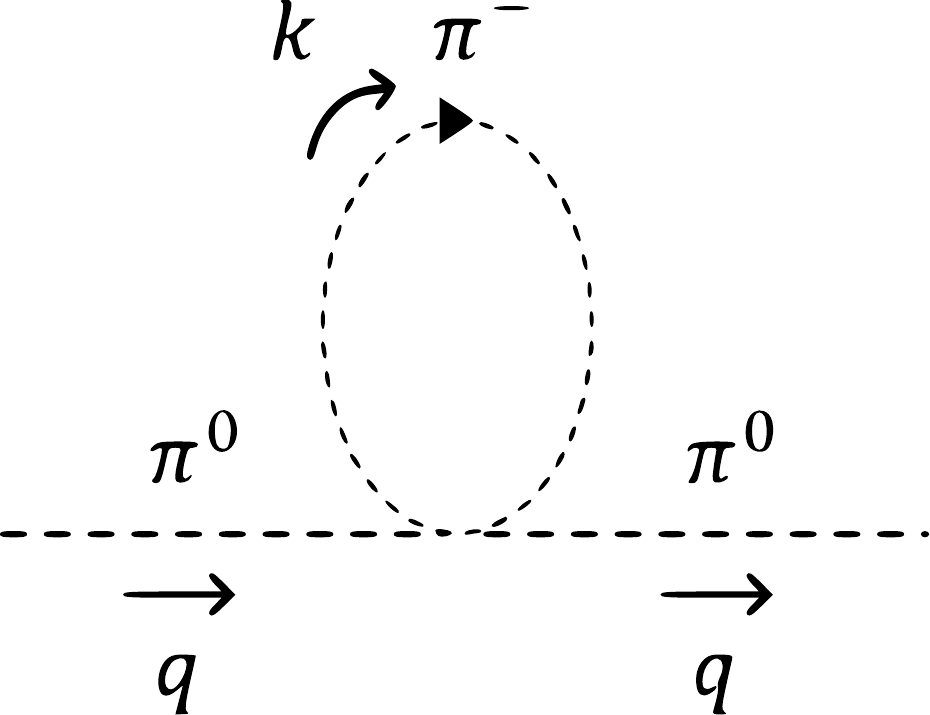}
    \caption{Feynman diagram corresponding t the one-loop contribution from charged pions to the neutral pion self-energy in the LSMq.}
    \label{self-energypions}
\end{figure}

\section{One-loop magnetic corrections} \label{sec4}

To compute the magnetic-field-induced modification to the neutral pion screening mass, the starting point is the equation defining its dispersion relation in the presence of the magnetic field, namely,
\begin{eqnarray}
  q_0^2-|\vec{q}|^2-m_0^2(B)-{\mbox{Re}}[\Pi_B]=0,
\label{findsol}
\end{eqnarray}
where $\Pi_B$ is the magnetic-field-dependent neutral pion self-energy that depends on the model couplings and masses. Notice that, for the calculation of the magnetic-field-induced modifications to the mass, only the real part of $\Pi_B$ contributes. On the other hand, the imaginary part would contribute to the magnetic-field-induced pion damping rate. The properties of this damping rate can also offer insights into the possible opening of magnetic field driven channels for particle process; however, for the purposes of the present work, we hereby concentrate exclusively on the magnetic-field-induced modifications of the (longitudinal screening) mass that are encoded in the real part of the self-energy.

The computation requires knowledge of each of the above-mentioned elements as functions of the field strength. To obtain the screening mass, we need to set $q_{0}=0$ in Eq.~(\ref{findsol}) and find positive solutions for the parameter $m_{sc}^2=-|\vec{q}|^2$. In the presence of a constant magnetic field, we have two kinds of solutions for $m_{sc}^2$: the longitudinal screening mass denoted by $m_{sc,\parallel}$, which is defined for the limit where $\vec{q}_\perp = 0$, and the transverse screening mass, denoted by $m_{sc,\perp}$, which is defined for the limit where $q_3=0$. Since we have chosen the direction of the magnetic field to point along the $z$ axis, $m_{sc,\parallel}^2=-q^{2}_{3}$, whereas $m_{sc,\perp}^2=-|\vec{q_{\perp}}|^2$. In what follows, we concentrate on the calculation of the longitudinal screening mass. We leave for a future work the computation of the transverse screening mass.

\begin{figure}[t]
    \includegraphics[scale=1.1]{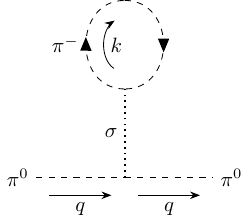}
    \caption{Feynman diagram corresponding to the tadpole contribution from charged pions and a sigma to the neutral pion self-energy in the LSMq.}
    \label{tadpole2}
\end{figure}
We first compute the neutral pion self-energy,
\begin{equation}
 \Pi^B=\sum_f\left(\Pi_{f\bar{f}}^B(q)+\Pi_f^B\right)+\Pi_{\pi^\pm}^B+\Pi_{\pi^0}+\Pi_{\sigma}.
 \label{totalselfenergy}
\end{equation}
The terms on the right-hand side of Eq.~(\ref{totalselfenergy}) are represented by the Feynman diagrams depicted in Figs.~\ref{self-energyquarks},~\ref{tadpole1},~\ref{self-energypions} and~\ref{tadpole2}, which contribute to the self-energy at one loop. The subindices represent the kind of particles in the loop and correspond to the quark-antiquark loop $\Pi_{f\bar{f}}^B$ depicted in Fig.~\ref{self-energyquarks}, the quark tadpole $\Pi_f^B$ depicted in Fig.~\ref{tadpole1}, the charged boson tadpoles $\Pi_{\pi^\pm}^B$, depicted in Figs.~\ref{self-energypions} and~\ref{tadpole2},
and the neutral boson tadpoles $\Pi_{\pi^0},\Pi_{\sigma}$. Notice that the diagrams with neutral bosons in the loop contribute only to vacuum renormalization and not to the magnetic properties of the system. To see this, recall that these fields contribute with terms represented by the regularized integrals that are computed using bare couplings and masses. When the propagator does not include magnetic field effects, the result of the regularized integral can be canceled by the introduction of suitable counter-terms. The upshot is that, when a given diagram is computed without field effects in the propagator, it does not contribute to the magnetic field modifications of particle properties. Therefore, hereafter we do not consider the effect of these diagrams for the description of the magnetic modifications of the pion self-energy. 

Since the contribution from the quark-antiquark loop is the only one that depends on the pion momentum, we first concentrate on the contribution from this diagram, for a single quark species. This is given explicitly by
\begin{eqnarray}
    \!\!\!\!\!\!-i\Pi^{B}_{f\bar{f}}(q)&\!\!\!=\!\!\!&-g^2\!\int \frac{d^{4}k}{(2\pi)^{4}}{\mbox{Tr}}[\gamma_{5}iS_{f}(k)\gamma_{5}iS_{f}(k+q)],
    \label{quarkloop}
\end{eqnarray}
Notice that, since both particles flow with the same charge around the loop, the Schwinger phase vanishes. According to the explicit computation in the Appendix, the fermion contribution to the pion self-energy is given by
\begin{eqnarray}
    \Pi^{B}_{f\bar{f}}(q)&=&-4g^{2}\frac{|q_{f}B|}{(4\pi)^{2}}\int_{0}^{1}dv\int_{0}^{\infty}du\nonumber\\
    &\times&\exp[-i\frac{q^{2}_{\perp}}{|q_{f}B|}\frac{\sin(|q_{f}B|u(1-v))\sin(|q_{f}B|uv)}{\sin(|q_{f}B|u)}]\nonumber\\
    &\times&e^{-iq^{2}_{3}uv(1-v)}e^{iq^{2}_{0}uv(1-v)}e^{-ium^{2}_{f}}e^{-u\epsilon}\nonumber\\
    &\times&\Bigg\{\frac{m^{2}_{f}}{\tan(|q_{f}B|u)}
    +\frac{|q_{f}B|} {\sin^{2}(|q_{f}B|u)}\nonumber\\ &\times&\Bigg(\frac{-q^{2}_{\perp}}{|q_{f}B|}\frac{\sin(|q_{f}B|u(1-v))\sin(|q_{f}B|uv)}{\sin(|q_{f}B|u)}-i\Bigg)\nonumber\\
    &+&\frac{1}{u\tan(|q_{f}B|u)}\bigg(\frac{1}{i}-uv(1-v)(q^{2}_{3}-q^{2}_{0})\Bigg)\Bigg\}\,.\nonumber\\
    \label{Self_Energy_3p}
\end{eqnarray}
where we have defined the variables
\begin{align}
     s&=u(1-v), \nonumber \\
     s^{\prime}&=uv\,.
\label{New_Variables}
\end{align}

To isolate the magnetic contribution in the pion self-energy, we need to work with the function $F(q^{2}_{0},q^{2}_{3},q^{2}_{\perp},|q_{f}B|,m_{f})$ defined as
\begin{equation}
F(q^{2}_{0},q^{2}_{3},q^{2}_{\perp},|q_{f}B|,m_{f})=\Pi^{B}_{f\bar{f}}-\lim_{q_{f}B\rightarrow0}\Pi^{B}_{f\bar{f}}\,,
\label{F}
\end{equation}
explicitly given by
\begin{widetext}
\begin{eqnarray}    F(q^{2}_{0},q^{2}_{3},q^{2}_{\perp},|q_{f}B|,m_{f})&=&-4g^{2}\frac{|q_{f}B|}{(4\pi)^{2}}\int_{0}^{1}dv\int_{0}^{\infty}due^{-u\epsilon}\nonumber\\ &\times&\Bigg\{
e^{-iX}\Bigg[\frac{m^{2}_{f}}{\tan(|q_{f}B|u)}-\frac{q^{2}_{\perp}\sin(|q_{f}B|u(1-v))\sin(|q_{f}B|uv)}{\sin^{3}(|q_{f}B|u)}-\frac{v(1-v)(q^{2}_{3}-q^{2}_{0})}{\tan(|q_{f}B|u)}\nonumber\\
&-&i\Bigg(\frac{|q_{f}B|}{\sin^{2}(|q_{f}B|u)}+\frac{1}{u\tan(|q_{f}B|u)}\Bigg)\Bigg]\nonumber\\ &-&e^{-iX_{0}}\Bigg[\frac{(m^{2}_{f}-v(1-v)(q^{2}_{3}+q^{2}_{\perp})+v(1-v)q^{2}_{0})u-2i}{|q_{f}B|u^2}\Bigg]\Bigg\}\,,
\label{f_function}
\end{eqnarray}
\end{widetext}
where $X$ and $X_{0}$ are defined as
\begin{align}
     X&=\frac{q^{2}_{\perp}}{|q_{f}B|}\frac{\sin(|q_{f}B|u(1-v))\sin(|q_{f}B|uv)}{\sin(|q_{f}B|u)}+q^{3}_{3}uv(1-v) \nonumber \\
     &-q^{2}_{0}uv(1-v)+m^{2}_{f}u,\nonumber\\
     X_{0}&=uv(1-v)(q^{2}_{\perp}+q^{2}_{3})-q^{2}_{0}uv(1-v)+m^{2}_{f}u\,. 
\label{XX0}
\end{align}
Hereafter, we concentrate on the computation of the longitudinal screening mass. For this purpose, we set $q^{2}_{\perp}=q^{2}_{0}=0$ in the function $F(q^{2}_{0},q^{2}_{3},q^{2}_{\perp},|q_{f}B|,m_{f})$ of Eq.~(\ref{f_function}), to obtain
\begin{equation}
\begin{split}
    &F(0,q^{2}_{3},0,|q_{f}B|,m_{f})=-4g^{2}\frac{|q_{f}B|}{(4\pi)^{2}}\int_{0}^{1}dv\int_{0}^{\infty}due^{-u\epsilon}\\&\times\Bigg\{e^{-iX}\Bigg[\frac{m^{2}_{f}-v(1-v)q^{2}_{3}}{\tan(|q_{f}B|u)}\\&-i\Bigg(\frac{|q_{f}B|}{\sin^{2}(|q_{f}B|u)}+\frac{1}{u\tan(|q_{f}B|u)}\Bigg)\Bigg]\\&-e^{-iX_{0}}\Bigg[\frac{m^{2}_{f}-v(1-v)q^{2}_{3}}{|q_{f}B|u}-\frac{2i}{|q_{f}B|u^{2}}\Bigg]\Bigg\}\,.
    \end{split}
    \label{f_function_2}
\end{equation}
In this case, $X$ and $X_{0}$ reduce to the same expression, which is explicitly given by
\begin{equation}
    X=X_{0}=uv(1-v)q^{2}_{3}+u m^{2}_{f} \equiv ua\,,
\end{equation}
with 
\begin{equation}
   a\equiv m^{2}_{f}+v(1-v)q^{2}_{3}\,. 
   \label{a}
\end{equation}
\begin{figure}[t]
    \centering
    \includegraphics[scale=0.60]{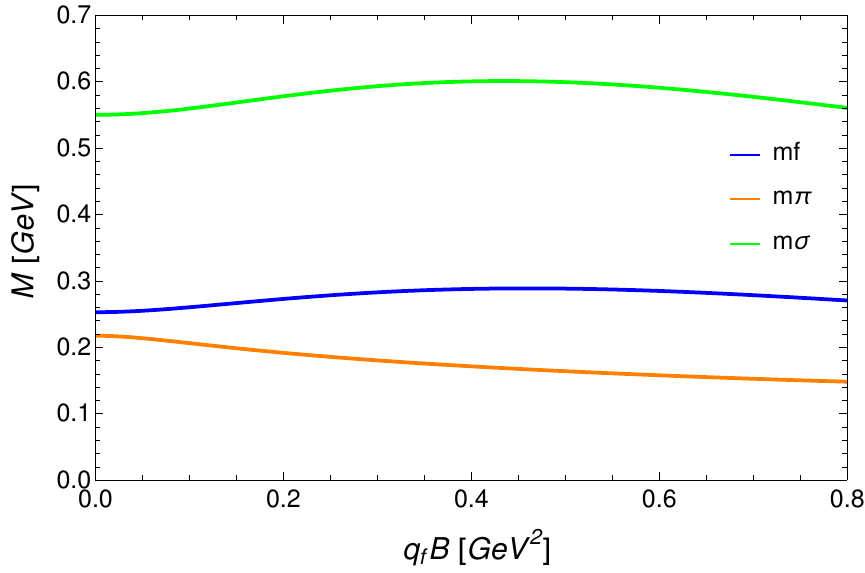}
    \caption{Masses of the quark (blue line), pion (orange line), and sigma (green line) as a function of the magnetic field following Ref.~\cite{nosso03}}
    \label{Masses}
\end{figure}
The real part of the $u$ integral in Eq.~(\ref{f_function_2}), which is needed to compute the screening mass, can be performed analytically (see the Appendix), with the result
\begin{equation}
\begin{split}
&\Re [F(0,q^{2}_{3},0,|q_{f}B|,m_{f})]=-\frac{4g^{2}}{(4\pi)^{2}}\int_{0}^{1}dv\Bigg\{\\&-2v(1-v)q^{2}_{3}\Bigg[A_{1}+A_{2}\Bigg]\\&+|q_{f}B|\Bigg[\frac{\epsilon\pi}{2|q_{f}B|}\!-\!\ln\!\Bigg(\sqrt{2\cosh\left(\frac{\epsilon\pi}{|q_{f}B|}\right)-2\cos\left(\frac{a\pi}{|q_{f}B|}\right)}\Bigg)\!\Bigg]\\&-\frac{|q_{f}B|}{\pi}\Bigg[\Re\Bigg(Li_{2}\left(e^{-(ia+\epsilon)\frac{\pi}{|q_{f}B|}}\right)\Bigg)\Bigg]\\&+|q_{f}B|\ln\left(\frac{a}{|q_{f}B|}\right)-a\ln\left(\frac{a}{2|q_{f}B|}\right)+a-|q_{f}B|\ln(4\pi)\\&+2|q_{f}B|\ln\Gamma\left(\frac{a}{2|q_{f}B|}\right)\Bigg\}\,,
\end{split}
\label{previoustoAB}
\end{equation}
where $A_{1}$ and $A_{2}$ are given by
\begin{align}
     A_{1}&=\Bigg[\frac{\pi}{2}\frac{\sin(\frac{a\pi}{|q_{f}B|})}{\cosh(\frac{\epsilon\pi}{|q_{f}B|})-\cos(\frac{a\pi}{|q_{f}B|})} \nonumber \\
     &-\tan^{-1}\Bigg(\frac{e^{\frac{\epsilon\pi}{|q_{f}B|}}\sin(\frac{a\pi}{|q_{f}B|})}{1-e^{\frac{\epsilon\pi}{|q_{f}B|}}\cos(\frac{a\pi}{|q_{f}B|})}\Bigg)\Bigg],\nonumber\\
     A_{2}&=-\frac{|q_{f}B|}{a}+\ln\left(\frac{a}{2|q_{f}B|} \right)-\psi^{(0)}\left(\frac{a}{2|q_{f}B|}\right)\,. 
\label{A_B}
\end{align}
Notice that the limits $q_fB\to 0$ and $\epsilon\to 0$ do not commute. Therefore, to check that Eq.~(\ref{previoustoAB}) goes to zero when $q_fB$ vanishes, the $\epsilon$ dependence has to be maintained. For finite and arbitrary values of $q_fB$, the integration over $v$ needs to be numerically performed for finite values of $\epsilon$. We have checked that the value of the integral converges after having performed the $v$ integration when we then take smaller values of $\epsilon$ so as to implement the limit $\epsilon\to 0$.
\begin{figure}[t]
    \centering
    \includegraphics[scale=0.28]{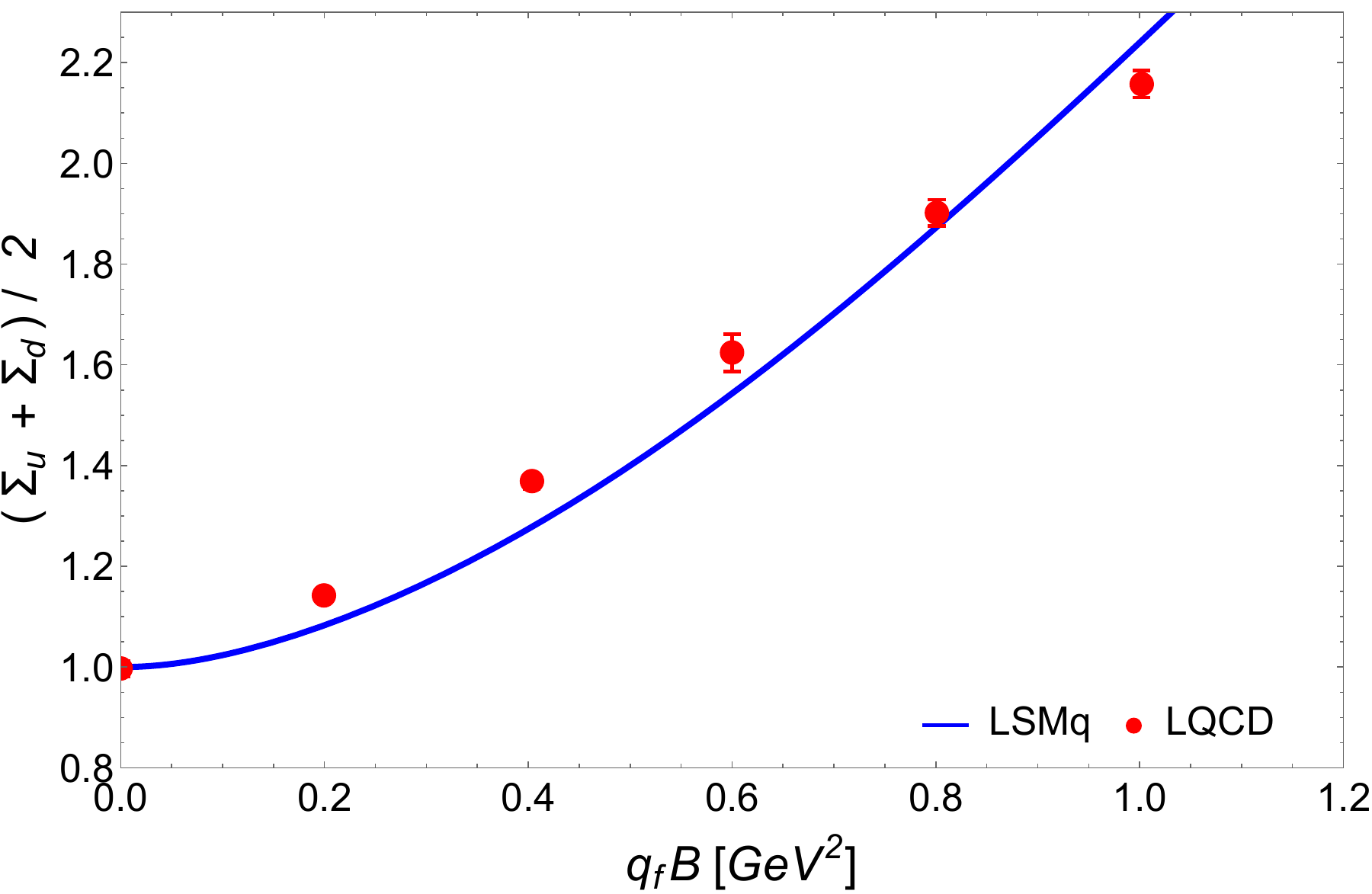}
    \caption{Average normalized condensate computed from the model compared to the results from Ref.~\cite{Bali}.}
    \label{Condensate}
\end{figure}
Notice that, to compute the magnetic modification of the screening mass, only the real part of the self-energy is required. However, the imaginary part of the self-energy is also an interesting and useful quantity that could be computed in the presence of a magnetic field, since this is directly linked to the magnetic field activation of decay channels that are otherwise not present in the absence of magnetic fields. For example, if as a consequence of field effects, a meson mass becomes larger than twice the quark mass, the meson decay into a quark-antiquark pair can be opened and this is signaled by a nonvanishing imaginary part of the meson self-energy. In the present work, no such channel is opened since the pion mass decreases and always remains smaller than twice the quark mass as a function of the field strength, and consequently, the magnetic-field-dependent imaginary part of the pion self-energy vanishes, as expected. This may not be the case were we to consider the $\sigma$ meson or when the combined thermomagnetic effects on meson masses are considered.

We now proceed to compute the charged boson loop contribution to the neutral pion self-energy. This includes the two tadpole diagrams shown in Figs.~\ref{self-energypions} and \ref{tadpole2} and can be written as 
\begin{equation}
    \Pi^{B}_{\pi^{\pm}}=\frac{8\lambda}{4}\left(1-\frac{\lambda f^{2}_{\pi}}{m^{2}_{\sigma}}\right)\Pi^{B}_{b}\,,
    \label{bosonicselfenergy}
\end{equation}
where $\Pi^{B}_{b}$ is the contribution to the neutral pion self-energy coming from Fig.~\ref{self-energypions}, which is calculated by
\begin{equation}
    -i\Pi^{B}_{b}=\int \frac{d^{4}k}{(2\pi)^{4}}D_{\pi^{\pm}}(k).
    \label{chargedself}
\end{equation}
Notice that, since the initial and final loop space-time points in the tadpole Feynman diagram coincide, the Schwinger phase also vanishes.
Substituting Eq.~(\ref{bosonpropagatormomentumspace}) into Eq.~(\ref{chargedself}) and integrating over the momentum and $s$ variables, we obtain, after subtracting the $B=0$ contribution,
\begin{equation}
\begin{split}
    \Pi^{B}_{\pi^{\pm}}&=\frac{m^{2}_{b}}{16\pi^{2}}\left[\ln\left(\frac{m^{2}_{b}}{2|q_{b}B|}\right)-1\right]+\frac{|q_{b}B|}{16\pi^{2}}\ln(2\pi)\\
    &-\frac{|q_{b}B|}{8\pi^{2}}\ln\left[\Gamma\left(\frac{1}{2}+\frac{m^{2}_{b}}{2|q_{b}B|}\right)\right]\,.
\end{split}
\label{tadpolereal}
\end{equation}
\begin{figure}[t]
    \centering
    \includegraphics[scale=0.59]{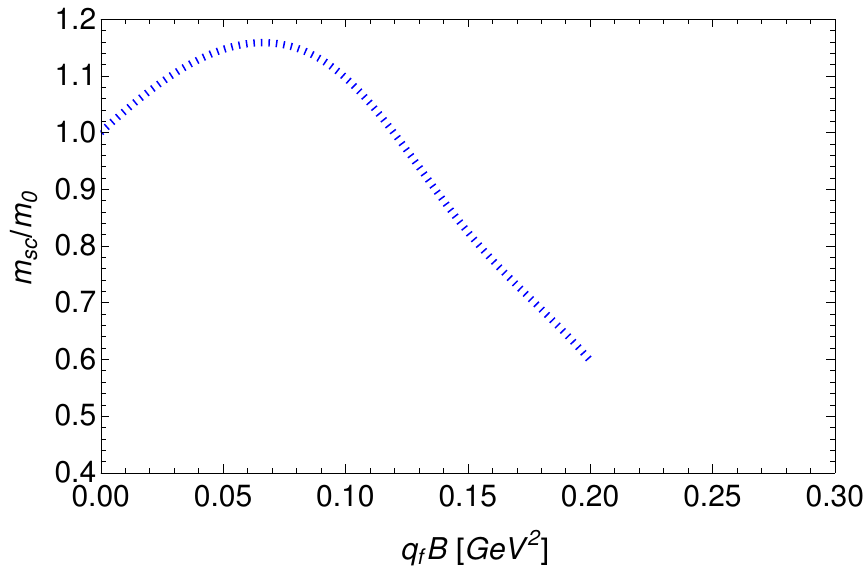}
    \caption{Neutral pion longitudinal screening mass as a function of the magnetic field strength normalized to the pion pole mass for $B=0$ computed using g=2.75. Solutions to the dispersion relation cease to exist beyond $q_fB\sim 0.2$ GeV$^2$.}
\label{figfixedgvacuum}
\end{figure}
Notice that, since
Eq.~(\ref{tadpolereal}) does not depend on the external momentum, it represents a purely real contribution. Therefore, the explicit expression for $\Pi^{B}_{\pi^{\pm}}$ in Eq.~(\ref{bosonicselfenergy}) is given by
\begin{equation}
\begin{split}
    \Pi^{B}_{\pi^{\pm}}&=\frac{8\lambda}{4}\left(1-\frac{\lambda f^{2}_{\pi}}{m^{2}_{\sigma}}\right)\Bigg\{\frac{m^{2}_{b}}{16\pi^{2}}\left[\ln\left(\frac{m^{2}_{b}}{2|q_{b}B|}\right)-1\right]\\&+\frac{|q_{b}B|}{16\pi^{2}}\ln(2\pi)-\frac{|q_{b}B|}{8\pi^{2}}\ln\left[\Gamma\left(\frac{1}{2}+\frac{m^{2}_{b}}{2|q_{b}B|}\right)\right]\Bigg\}\,.
\end{split}
\end{equation}
Finally, for the contribution of the quark tadpole $\Pi^{B}_{f}$ shown in Fig.~\ref{tadpole1}, we have
\begin{equation}
    -i\Pi^{B}_{f}=2\lambda vg\int\frac{d^{4}k}{(2\pi)^{4}}{\mbox{Tr}}[iS(p)]\,.
    \label{fermiontadpole}
\end{equation}
Substituting Eq.~(\ref{fermionpropagatormomentumspace}) into Eq.~(\ref{fermiontadpole}), then performing a Wick rotation to Euclidian space and finally integrating over the momentum variables, we obtain, after subtracting the $ B=0$ contribution,
\begin{equation}
    \Pi^{B}_{f}=\frac{\lambda gm_{f}f_{\pi}}{8\pi^{2}m^{2}_{\sigma}}\int_{0}^{\infty}ds\frac{e^{-s m^{2}_{f}}}{s^{2}}\left[\frac{|q_{f}B|}{\tanh(|q_{f}B|s)}-1\right]\,.
\label{tadpolefermion}
\end{equation}
Notice that Eq.~(\ref{tadpolefermion}) corresponds also to a purely real contribution.
\begin{figure}[t]
    \centering
    \includegraphics[scale=0.55]{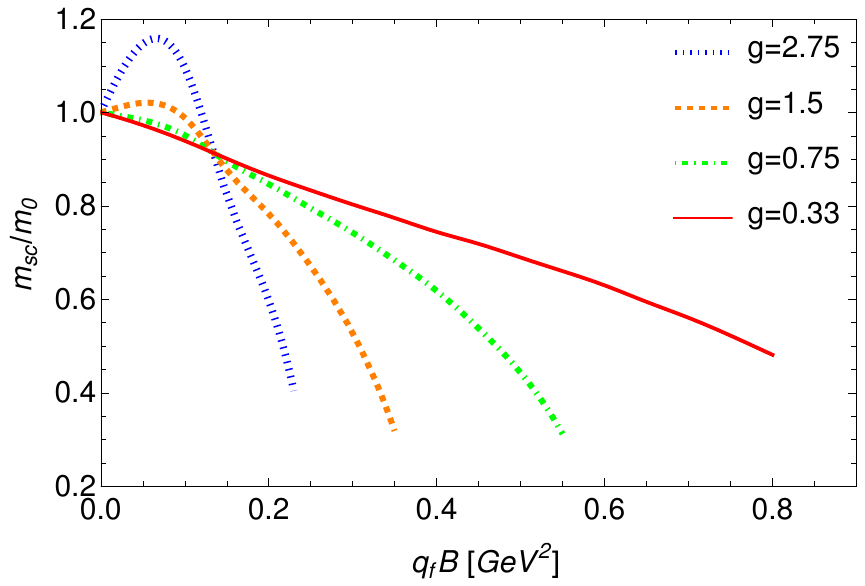}
    \caption{Neutral pion longitudinal screening mass as a function of the magnetic field strength normalized to the pion pole mass for $B=0$: $g=2.75$, (dotted), $g=1.5$ (dashed), $g=0.75$ (dashed-dot), and $g=0.33$ (solid). Solutions to the dispersion relation cease to exist beyond $q_fB\sim 0.2, 0.35, 0.55$ and 0.8 GeV$^2$, respectively, and the range where solutions exist increases as the coupling decreases.}
    \label{figseveralg}
\end{figure}

\section{Magnetic modification to the neutral pion mass} \label{sec5}

With all these elements at hand, we can now find the magnetic-field-dependent 
longitudinal screening mass for the neutral pion from the dispersion relation (\ref{findsol}) by setting $q^{2}_{\perp}=q^{2}_{0}=0$. Since we are pursuing the purely magnetic field effects, we also subtract the $B=0$ contribution, which amounts to subtracting the vacuum contribution, namely, 
\begin{equation}
-q^{2}_{3}=m_{0}^{2}(B)+\Re \mathbb{F}(q^{2}_0=0,q^{3}_{3},q^{2}_{\perp}=0,|q_{f}B|)
\label{selfconsistenteq}\,,
\end{equation}
where $\mathbb{F}$ is defined in correspondence with Eq.~(\ref{F}) and accounting for all relevant diagrams, that is,
\begin{equation}
\mathbb{F}(q^{2}_{0},q^{2}_{3},q^{2}_{\perp},|q_{f}B|)=\Pi^{B}-\lim_{q_{f}B\rightarrow0}\Pi^{B}\,.
\label{F2}
\end{equation}
The longitudinal screening mass is  obtained finding solutions for $m_{sc,\parallel}^{2}\equiv -q^{2}_{3}$, for different values of the field strength. In anticipation of the results, we point out that, in order to make a reasonable description of the behavior of the screening mass with the field strength, we need to account for the magnetic field dependence of the different particles involved in the self-energy, as well as of the couplings. In this sense, the full-fledged description of the problem therefore requires a self-consistent treatment, whereby all self-energies of the particles subject to the influence of the magnetic field depend on each other through the field dependence of their masses. However, for our purposes, here we set the problem in a simpler manner. We borrow results for the  magnetic field dependence of the pion, $\sigma$ and quark pole masses, which are inputs to compute the magnetic corrections to the neutral pion screening mass. We have taken as input the pole pion mass $m_{0}(B)$, the quark mass $m_{f}(B)$, and the $\sigma$ mass $m_{\sigma}(B)$ as functions of the magnetic field from Ref.~\cite{nosso03}. Figure~\ref{Masses} shows the magnetic field dependence of the input masses. To have a direct comparison with LQCD results of Ref.~\cite{Ding-2022}, hereby we use a vacuum value of $m_0(B=0)=220$ MeV for the pion mass, $m_{f}(B=0)=252$ MeV for the quark mass, and $m_{\sigma}(B=0)=550$ MeV for the $\sigma$ mass. In principle, the magnetic mass dependence we use is rigorously valid for $eB \leq 0.4$ GeV$^2$, which is the upper limit for the cutoff for the NJL calculation of Ref. [72]. Hence, the mass values for large magnetic fields should be considered as extrapolations, as they provide only a qualitative behavior in this limit. As we show, the magnetic dependence of these masses turns out to be a key ingredient that allows a good  description of the behavior of the longitudinal screening mass found by LQCD and for NJL model-based calculations~\cite{Ding-2022,Sheng}. 

Before proceeding to the analysis of the screening mass, we first test whether the model can be used to describe the LQCD average condensate as a function of the field strength. Figure~\ref{Condensate} shows this quantity taken from Ref.~\cite{Bali} compared to our model calculation, using the same magnetic field dependence of the quark mass that we use as input to compute the neutral pion screening mass. The model calculation provides a reasonable description of the LQCD data.
\begin{figure}[t]
    \centering
    \includegraphics[scale=0.55]{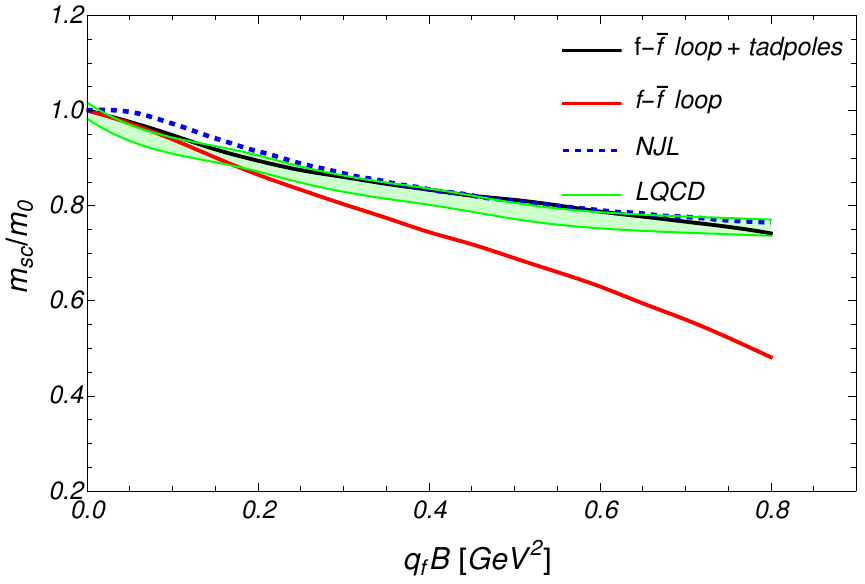}
    \caption{Neutral pion longitudinal screening mass as a function of the magnetic field strength normalized to the pion pole mass for $B=0$ including all the contributions to the self-energy computed with $g=0.33$ and $\lambda=2.5$ compared to the  NJL results from Ref.~\cite{Sheng} and to an interpolation of the data for the LQCD results from Ref.~\cite{Ding-2022} for $T=17$ MeV. The green shadow represents the error in the LQCD calculations from Ref.~\cite{Ding-2022}. For comparison we also show the case where only the fermion-antifermion loop is considered, computed with $g=0.33$.}
    \label{figfinalresult}
\end{figure}

As discussed in Sec.~\ref{sec4}, the neutral pion self-energy is described by the two couplings $g$ and $\lambda$; the former enters in the calculation of the fermion contribution to the self-energy, depicted in Fig.~\ref{self-energyquarks}, whereas the latter enters in the contribution of the tadpole diagrams of Figs.~\ref{self-energypions} and~\ref{tadpole2}. Also, a combination of both couplings enters in the computation of the tadpole diagram in Fig.~\ref{tadpole1}. In vacuum, these parameters have to obey the following constraints imposed by the model and that are derived from Eq.~(\ref{masses}):
\begin{align}
     g&=m_{f}/f_\pi\,, \nonumber \\
     \lambda&=\frac{m^{2}_{\sigma}-m^{2}_{0}}{2f_\pi^{2}}\,,
\label{constraints}
\end{align}
where, in account of the partially conserved axial current statement, we  identify the vacuum expectation value $v_0$ with $f_\pi$, the pion decay constant.
Substituting the values of the masses in Eq.~(\ref{constraints}), we obtain $g\sim 2.75$ and $\lambda\sim 15$.

We now use the aforementioned parameters in Eq.~(\ref{selfconsistenteq}) to find the screening mass for the neutral pion. The results are shown in Fig.~\ref{figfixedgvacuum} as the ratio  $m_{sc,\parallel}/m_{0}$. Hereafter, for the calculations, we sum over the two light quark flavor charges taking $|q_u|=2/3$ and $|q_d|=1/3$. Notice that, with this choice, the behavior of the screening mass does not resemble the findings of LQCD, nor those of NJL. Furthermore, the solutions to the dispersion relation equation cease to exists for an intermediate value of the field strength.

Motivated by the results of Ref.~\cite{nosso03}, which point out to a fast decrease of the NJL coupling as a function of the magnetic field, we first study the consequences of using a lower value of the $g$ coupling to explore the effects for the  $m_{sc,\parallel}/m_{0}$ ratio. The results are shown in Fig.~\ref{figseveralg}. Notice that the effect of decreasing $g$ is to increase the range of solutions for $m_{sc,\parallel}$ as a function of $q_fB$, producing results closer to those of the NJL and LQCD ones. We find that the choice $g=0.33$, which corresponds to the solid line plot in Fig.~\ref{figseveralg}, already provides a good description of the NJL and LQCD findings. Finally, we add the contribution from the tadpoles shown in Figs.~\ref{tadpole1} and \ref{tadpole2}. Here, we naturally choose the best parameter already determined for $g$, that is, $g=0.33$ from Fig.~\ref{figseveralg}, and use it as a starting point to then add the tadpole contributions. 
The results are shown in Fig.~\ref{figfinalresult} for the choice of parameters $g=0.33$ and $\lambda=2.5$. Here, we compare our findings with the results for the screening masses reported in Ref.~\cite{Sheng} for the NJL model and also with the LQCD results in Ref.~\cite{Ding-2022} for $T=17$ MeV. Notice that the NJL results are reported for $T=0$, just as ours, however, the results from LQCD are calculated for finite temperature. We thus compare with the smallest temperature reported, which corresponds to $T=17$ MeV. The results are consistent with the findings in Refs.~\cite{AHHFZ, Ayala1} for the magnetic field dependence of the pole pion mass in the large field limit. We emphasize that a good description for the LQCD and NJL results for the neutral pion parallel screening mass can be achieved only when the couplings are taken to be about one order of magnitude smaller than their vacuum values. We have refrained from parametrizing the magnetic field dependence of these couplings, but instead highlight that their decrease happens soon after the magnetic field starts growing from zero.

\section{Summary and conclusions}\label{sec6}

In this work we studied the magnetic-field-induced modifications on the longitudinal screening mass of the neutral pion at one-loop level using the LSMq. The effects of the magnetic field are introduced in the neutral pion self-energy which is made out of several terms stemming from the contribution from the $\sigma$ as well as from the charged particles of the model to the loop corrections. We found that, in order to obtain a reasonable description for the behavior of the longitudinal screening mass with the field strength, the magnetic field dependence of the particle masses, as well as of the couplings, needs to be taken into account. Moreover, for the calculation to reproduce the corresponding results from LQCD and NJL, the couplings $g$ and $\lambda$ need to decrease fast enough (within a magnetic field interval $\simeq 0.1$ GeV$^2$ from $B=0$) to then reach constant and small values with the field strength. This result is in agreement with the findings of Refs.~\cite{AHHFZ,Ayala1}. The results illustrate the need to account for the backreaction of the magnetic field dependence of the rest of the particle species in the model. This could be achieved by a complete self-consistent treatment of the problem. However, this represents a highly involved procedure, requiring the simultaneous solution, at a given perturbative order, first of the set of coupled equations that govern the behavior of the pole masses, together with the couplings, to then use these as inputs for the coupled set of equations that yield the screening masses. Although this procedure can be in principle implemented, in this work we have taken the more modest approach that makes use of the magnetic field behavior of the particle masses found in Ref.~\cite{Avancinipole1}. In this sense, we  believe that this work provides further evidence of the need to consider mutually dependent magnetic-field-dependent masses and couplings in effective model calculations to achieve better insight into the properties of strongly interacting systems subject to the effects of magnetic fields. The results are obtained using a method to analytically carry out the calculation of the quark-antiquark contribution to the neutral pion longitudinal screening mass up to the last integral. The method is valid for arbitrary field strengths, but cannot be directly applied to the case of the transverse screening mass, for which the pole contributions need to be handled in a different manner. We are currently working on this calculation, and the results, together with thermal effects, will soon be reported elsewhere.

\section*{Acknowledgements}

Support for this work was received in part by UNAM-PAPIIT-IG100322 and by Consejo Nacional de Humanidades, Ciencia y Tecnolog\'ia grant numbers CF-2023-G-433, A1-S-7655 and A1-S-16215. J. Rend\'on acknowledges support from the program estancias
posdoctorales por M\'exico of CONAHCyT.  R. Zamora acknowledges support from FONDECYT (Chile) under grant No. 1200483. 
C. Villavicencio acknowledges support from FONDECYT (Chile) under grants 1190192 and 1220035.
This work is partially supported by Conselho Nacional de Desenvolvimento Cient\'ifico e Tecnol\'ogico  (CNPq), Grant No. 304758/2017-5 (R. L. S. F); Funda\c{c}\~ao de Amparo \`a Pesquisa do Estado do Rio Grande do Sul (FAPERGS), Grants Nos. 19/2551- 0000690-0 and 19/2551-0001948-3 (R. L. S. F.) and also part of the project Instituto Nacional de Ci\^encia 
e Tecnologia - F\'isica Nuclear e Aplica\c{c}\~oes (INCT - FNA), Grant No. 464898/2014-5 (R. L. S. F).
\begin{appendix}
\section{Neutral pion self-energy calculation}\label{sec7}
Consider the quark-antiquark loop depicted in Fig.~\ref{self-energyquarks}. The loop can be made of either quarks $u$ or $d$. For a quark of flavor $f$, its contribution to the neutral pion self-energy is given by Eq.~(\ref{quarkloop}),
where we used the fact that the Schwinger phase vanishes. To proceed with the calculation of $-i\Pi^{B}_{f\bar{f}}(q)$, we need to insert the fermion propagator, whose explicit form is given by Eq.~(\ref{fermionpropagatormomentumspace}).
We find that only five of the traces survive. The surviving terms are given by
\begin{widetext}
\begin{eqnarray}
  Tr[\gamma^{5}m^{2}_{f}\cos(|q_{f}B|s)\gamma^{5}\cos(|q_{f}B| s^{\prime})] & = & 4m^{2}_{f}\cos(|q_{f}B|s)\cos(|q_{f}B|s^{\prime})\,, \label{Trace_1}\nonumber\\
  Tr[\gamma^{5}\slashed{k}_{\parallel}\cos(|q_{f}B|s)\gamma^{5}(\slashed{k}_{\parallel}+\slashed{q}_{\parallel})\cos(|q_{f}B|s^{\prime})]  &= & -4\cos(|q_{f}B|s)\cos(|q_{f}B|s^{\prime})k_{\parallel}\cdot(k_{\parallel}+q_{\parallel})\,, \label{Trace2}\nonumber\\
 Tr[\gamma^{5}m_{f}(\gamma^{1}\gamma^{2})\sin(|q_{f}B|s)\gamma^{5}m_{f}(\gamma^{1}\gamma^{2})\sin(|q_{f}B|s^{\prime})]  &= & -4m^{2}_{f}\sin(|q_{f}B|s)\sin(|q_{f}B|s^{\prime})\,,\label{Trace_3}\nonumber\\
 Tr[\gamma^{5}\slashed{k}_{\parallel}(\gamma^{1}\gamma^{2})\sin(|q_{f}B|s)\gamma^{5}(\slashed{k}_{\parallel}+\slashed{q}_{\parallel})(\gamma^{1}\gamma^{2})\sin(|q_{f}B|s^{\prime})]  & = & 4\sin(|q_{f}B|s)\sin(|q_{f}B|s^{\prime})k_{\parallel}\cdot(k_{\parallel}+q_{\parallel})\,,\label{Trace_4}\nonumber \\
 Tr\left[-\frac{\gamma^{5}\slashed{k}_{\perp}}{\cos(|q_{f}B|s)}(-\gamma^{5})\frac{(\slashed{k}_{\perp}+\slashed{q}_{\perp})}{\cos(|q_{f}B|s^{\prime})}\right] & = & \frac{4\vec{k}_{\perp}\cdot(\vec{k}_{\perp}+\vec{q}_{\perp})}{\cos(|q_{f}B|s)\cos(|q_{f}B|s^{\prime})}\,.
 \label{traces}
\end{eqnarray}
Substituting the values of these traces in the fermion contribution to the neutral pion self-energy, we obtain
\begin{equation}
\begin{split}
-i\Pi^{B}_{f\bar{f}}(q)=&-4g^{2}\int_{0}^{\infty}\int_{0}^{\infty}\frac{ds ds^{\prime}}{\cos(|q_{f}B|s)\cos(|q_{f}B|s^{\prime})}\int \frac{d^{4}k}{(2\pi)^{4}}e^{is\left[{k}^{2}_{\parallel}-k^{2}_{\perp}\frac{\tan(|q_{f}B|s)}{|q_{f}B|s}-m^{2}_{f}+i\epsilon\right]}\\&\times e^{is^{\prime}\left[({k}_{\parallel}+q_{\parallel})^{2}-(k_{\perp}+q_{\perp})^{2}\frac{\tan(|q_{f}B|s^{\prime})}{|q_{f}B|s^{\prime}}-m^{2}_{f}+i\epsilon\right]}\\&\times \left\{[\cos(|q_{f}B|(s+s^{\prime}))][m^{2}_{f}-{k}_{\parallel}\cdot({k}_{\parallel}+q_{\parallel})]+\frac{\vec{k}_{\perp}\cdot (\vec{k}_{\perp}+\vec{q}_{\perp})}{\cos(|q_{f}B|s)\cos(|q_{f}B|s^{\prime})}\right\}\,.
\end{split}
\label{Self_Energy_2}
\end{equation}
We proceed first to integrate over loop momentum components perpendicular to the magnetic field. The form of this integral is given by
\begin{equation}
I_{\perp}=e^{-\frac{b^{'}\vec{q}^{2}_{\perp}}{2}}\int_{-\infty}^{\infty}\frac{d^{2}\vec{k}_{\perp}}{(2\pi)^{2}}e^{-(a\vec{k}^{2}_{\perp}+b^{'}\vec{k}_{\perp}\cdot \vec{q}_{\perp})}\times\left[A+B\vec{k}_{\perp}\cdot (\vec{k}_{\perp}+\vec{q}_{\perp})\right]\,,
\end{equation}
where
\begin{subequations}
 \begin{align}
  a  = & \frac{i}{|q_{f}B|}(\tan(|q_{f}B|s)+\tan(|q_{f}B|s^{\prime}))\,, \\
  b^{\prime}  = &  \frac{2i\tan(|q_{f}B|s^{\prime})}{|q_{f}B|}\,,\\
 A  = & \cos(|q_{f}B|(s+s^{\prime}))(m^{2}_{f}-{k}_{0}(k_{0}+q_{0})+k_{3}(k_{3}+q_{3}))\,,\\
 B = & \frac{1}{\cos(|q_{f}B|s)\cos(|q_{f}B|s^{\prime})}\,.
\end{align}
\label{parameters}
\end{subequations}
The integral is trivially performed first by completing the square in the argument of the exponential and then using the known expressions for Gaussian integrals
\begin{subequations}
 \begin{align}
  \int_{-\infty}^{\infty}dxe^{-ax^{2}}  = & \sqrt{\frac{\pi}{a}}\,, \\
  \int_{-\infty}^{\infty}dx x^{2}e^{-ax^{2}}  = &  \sqrt{\frac{\pi}{a}}\frac{1}{2a}\,.
 \end{align}
 \label{Gaussian_Integrals}
\end{subequations}
The result for $I_{\perp}$ is 
\begin{equation}
I_{\perp}=\frac{e^{-\frac{bb^{\prime}}{4a}\vec{p}^{2}_{\perp}}B}{4\pi a}\left[A-\frac{B}{a}\left(\frac{bb^{\prime}}{4a}\vec{p}^{2}_{\perp}-1\right)\right]\,,
\label{Perpendicular_Integral}
\end{equation}
where, when completing the square, we found it convenient to introduce the parameter $b$ defined in terms of Eqs.~(\ref{parameters}) as
\begin{equation}
    b=2a-b^{\prime}=\frac{2i\tan(|q_{f}B|s)}{|q_{f}B|}\,.
\end{equation}
\end{widetext}
We now proceed to the calculation of the longitudinal momentum integral, whose explicit form is given by
\begin{equation}
I_{\parallel}=\int_{-\infty}^{\infty}\frac{dk_{3}}{2\pi}e^{-is k^{2}_{3}}e^{-is^{\prime}(k_{3}+q_{3})^{2}}\left\{\alpha+\beta k_{3}(k_{3}+q_{3})\right\}\,,
\end{equation}
with
\begin{subequations}
 \begin{align}
  \alpha  = & \frac{1}{i}\frac{m^{2}_{f}-k_{0}(k_{0}+q_{0})}{\tan(|q_{f}B|(s+s^{\prime}))}+\frac{|q_{f}B|}{\sin^{2}(|q_{f}B|(s+s^{\prime}))}\\&\times\left(\frac{i\vec{q}^{2}_{\perp}}{|q_{f}B|}\frac{\sin(|q_{f}B|s)\sin(|q_{f}B|s^{\prime})}{\sin(|q_{f}B|(s+s^{\prime}))}-1\right)\nonumber\,, \\
  \beta  = &  \frac{1}{i\tan(|q_{f}B|(s+s^{\prime}))}\,.
\end{align}
\end{subequations}
The integral $I_{\parallel}$ is also found with the help of the Gaussian integrals in Eq.~(\ref{Gaussian_Integrals}). The result is 
\begin{equation}
I_{\parallel}=\frac{e^{-iq^{2}_{3}\left(\frac{ss^{\prime}}{s+s^{\prime}}\right)}}{2\sqrt{\pi}(i(s+s^{\prime}))^{1/2}}\left\{\alpha+\frac{\beta}{s+s^{\prime}}\left(\frac{1}{2i}-\frac{ss^{\prime}}{s+s^{\prime}}q^{2}_{3}\right)\right\}\,.
\label{Parallel_Integral_Complete}
\end{equation}
Finally, for the integral over the zeroth momentum component, we have
\begin{equation}
I_{0}=\int_{-\infty}^{\infty}\frac{dk_{0}}{2\pi}e^{is k^{2}_{0}}e^{is^{\prime}(k_{0}+q_{0})^{2}}\left\{\mathcal{A}+\mathcal{B}k_{0}(k_{0}+q_{0})\right\}\,,
\end{equation}
where the $\mathcal{A}$ and $\mathcal{B}$ constants are defined as
\begin{eqnarray}
    \mathcal{A}&=&\frac{1}{i}\frac{m^{2}_{f}}{\tan(|q_{f}B|(s+s^{\prime}))}\nonumber\\&+&\frac{|q_{f}B|}{\sin^{2}(|q_{f}B|(s+s^{\prime}))}\nonumber\\&\times&\left(\frac{i\vec{q}^{2}_{\perp}}{|q_{f}B|}\frac{\sin(|q_{f}B|s)\sin(|q_{f}B|s^{\prime})}{\sin(|q_{f}B|(s+s^{\prime}))}-1\right)\nonumber\\
    &+&\frac{1}{i(s+s^{\prime})\tan(|q_{f}B|(s+s^{\prime}))}\left(\frac{1}{2i}-\frac{ss^{\prime}}{s+s^{\prime}}q^{2}_{3}\right),\nonumber\\
   \mathcal{B}&=&-\frac{1}{i\tan(|q_{f}B|(s+s^{\prime}))}\,.
   \end{eqnarray}
Using again the Gaussian integrals in Eq.~(\ref{Gaussian_Integrals}), the result for $I_{0}$ is given by
\begin{eqnarray}
I_{0}&=&\frac{e^{iq^{2}_{0}\left(\frac{ss^{\prime}}{s+s^{\prime}}\right)}}{2\sqrt{\pi}}\left(\frac{i}{s+s^{\prime}}\right)^{1/2}\nonumber\\
&\times&\left\{\mathcal{A}+\frac{\mathcal{B}}{s+s^{\prime}}\left(\frac{i}{2}-\frac{ss^{\prime}}{s+s^{\prime}}q^{2}_{0}\right)\right\}\,.
\label{Zero_Integral_Complete}
\end{eqnarray}
Substituting Eqs.~(\ref{Perpendicular_Integral}), (\ref{Parallel_Integral_Complete}), and (\ref{Zero_Integral_Complete}) into Eq.~(\ref{Self_Energy_2}), and making the change of variables $s=u(1-v)$ and $s^{\prime}=uv$, we get Eq.~(\ref{Self_Energy_3p}) that we hereby reproduce,
\begin{eqnarray}
    \Pi^{B}_{f\bar{f}}(q)&=&-4g^{2}\frac{|q_{f}B|}{(4\pi)^{2}}\int_{0}^{1}dv\int_{0}^{\infty}du\nonumber\\
    &\times&\exp[-i\frac{q^{2}_{\perp}}{|q_{f}B|}\frac{\sin(|q_{f}B|u(1-v))\sin(|q_{f}B|uv)}{\sin(|q_{f}B|u)}]\nonumber\\
    &\times&e^{-iq^{2}_{3}uv(1-v)}e^{iq^{2}_{0}uv(1-v)}e^{-ium^{2}_{f}}e^{-u\epsilon}\nonumber\\
    &\times&\Bigg\{\frac{m^{2}_{f}}{\tan(|q_{f}B|u)}
    +\frac{|q_{f}B|} {\sin^{2}(|q_{f}B|u)}\nonumber\\ &\times&\Bigg(\frac{-q^{2}_{\perp}}{|q_{f}B|}\frac{\sin(|q_{f}B|u(1-v))\sin(|q_{f}B|uv)}{\sin(|q_{f}B|u)}-i\Bigg)\nonumber\\
    &+&\frac{1}{u\tan(|q_{f}B|u)}\bigg(\frac{1}{i}-uv(1-v)(q^{2}_{3}-q^{2}_{0})\Bigg)\Bigg\}\,.\nonumber\\
    \label{Self_Energy_3}
\end{eqnarray}
As discussed in section~\ref{sec4}, to isolate the magnetic dependence in $\Pi^{B}_{f\bar{f}}(q)$, we need to subtract the self -energy evaluated at $B=0$ from the full pion self-energy. Therefore, we are interested in the function
\begin{equation}
F(q^{2}_{0},q^{2}_{3},q^{2}_{\perp},|q_{f}B|,m_{f})=\Pi^{B}_{f\bar{f}}-\lim_{q_{f}B\rightarrow0}\Pi^{B}_{f\bar{f}}\,.
\label{F_Function}
\end{equation}
Substituting Eq.~(\ref{Self_Energy_3}) into Eq.~(\ref{F_Function}), we obtain Eq.~(\ref{f_function}),
with $X$ and $X_{0}$ defined in Eq.~(\ref{XX0}).

To obtain the longitudinal screening mass, we need to set $q^{2}_{0}=q^{2}_{\perp}=0$ in Eq.~(\ref{f_function}), thus obtaining Eq.~(\ref{f_function_2}) that we show here explicitly for convenience,
\begin{equation}
\begin{split}
    &F(0,q^{2}_{3},0,|q_{f}B|,m_{f})=-4g^{2}\frac{|q_{f}B|}{(4\pi)^{2}}\int_{0}^{1}dv\int_{0}^{\infty}due^{-u\epsilon}\\&\times\Bigg\{e^{-iX}\Bigg[\frac{m^{2}_{f}-v(1-v)q^{2}_{3}}{\tan(|q_{f}B|u)}\\&-i\Bigg(\frac{|q_{f}B|}{\sin^{2}(|q_{f}B|u)}+\frac{1}{u\tan(|q_{f}B|u)}\Bigg)\Bigg]\\&-e^{-iX_{0}}\Bigg[\frac{m^{2}_{f}-v(1-v)q^{2}_{3}}{|q_{f}B|u}-\frac{2i}{|q_{f}B|u^{2}}\Bigg]\Bigg\}\,.
    \end{split}
    \label{F_Function_2}
\end{equation}
Equation~(\ref{F_Function_2}) can be conveniently written in the compact form
\begin{equation}
F(0,q^{2}_{3},0,|q_{f}B|,m_{f})\equiv-\frac{4g^{2}}{(4\pi)^{2}}\mathcal{F}(q^{2}_{3},|q_{f}B|,m_{f})\,,
\end{equation}
where
\begin{equation}
\begin{split}
\mathcal{F}(q^{2}_{3},|q_{f}B|,m_{f})\!&\!\equiv |q_{f}B|\int_{0}^{1}\!\!dv\int_{0}^{\infty}\!\!duG(u,v,q^{2}_{3},|q_{f}B|,m_{f}),
\end{split}
\end{equation}
with
\begin{equation}
\begin{split}
G(u,v,q^{2}_{3},|q_{f}B|,m_{f})&\equiv[m^{2}_{f}-v(1-v)q^{2}_{3}]e^{-u\epsilon}e^{-iau}\\&\times\left[\cot(|q_{f}B|u)-\frac{1}{|q_{f}B|u}\right]\\&-ie^{-u\epsilon}e^{-iau}\Bigg[|q_{f}B|\csc^{2}(|q_{f}B|u)\\&+\frac{\cot(|q_{f}B|u)}{u}-\frac{2}{|q_{f}B|u^{2}}\Bigg]\\&\equiv G_{1}-iG_{2}\,,
\end{split}
\end{equation}
where $a$ has been defined in Eq.~(\ref{a}).
Let us first study the part of the integral in $F(0,q^{2}_{3},0,|q_{f}B|,m_{f})$ that comes from the $G_{1}$ term, namely,
\begin{equation}
\begin{split}
I_{G_{1}}=&-\frac{4g^{2}}{(4\pi)^{2}}|q_{f}B|\int_{0}^{1}dv\int_{0}^{\infty}due^{-u\epsilon}(m^{2}_{f}-v(1-v)q^{2}_{3})\\&\times e^{-iau}\left(\cot(|q_{f}B|u)-\frac{1}{|q_{f}B|u}\right)\,.
\end{split}
\end{equation}
\begin{figure}[t]
    \centering
    \includegraphics[scale=0.33]{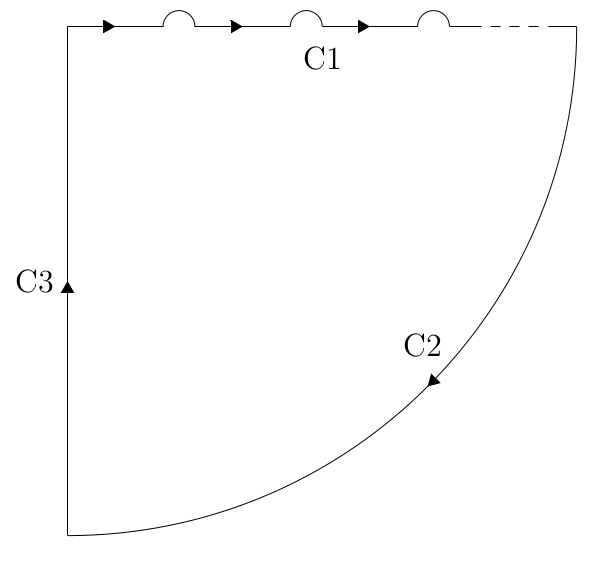}
    \caption{Contour of integration for Eq.~(\ref{I_Integral_Closed})}
    \label{figcontour}
\end{figure}
For the moment, let us focus only on the $u$ integral,
\begin{equation}
I\equiv\int_{0}^{\infty}due^{-u\epsilon}e^{-iau}\left(\cot(|q_{f}B|u)-\frac{1}{|q_{f}B|u}\right)\,.
\label{I_Integral}
\end{equation}
Since the poles of $\cot(|q_{f}B|u)$ lie along the real axis, we should evaluate Eq.~(\ref{I_Integral}) using the principal value prescription. Also, we promote the integral to the complex plane using the quarter circle contour shown in Fig.~\ref{figcontour}. Thus, we now focus in the contour integral,
\begin{equation}
I_{C}=\oint_{C}due^{-u\epsilon}e^{-iau}\left(\cot(|q_{f}B|u)-\frac{1}{|q_{f}B|u}\right)\,,
\label{I_Integral_Closed}
\end{equation}
where $C=C_{1}\cup C_{2}\cup C_{3}$, as shown in Fig.~\ref{figcontour}.
It is convenient to make the change of variables $u^{\prime}=|q_{f}B|u$ in our expression for $I_{C}$ so that
\begin{equation}
I_{C}=\oint_{C}\frac{du^{\prime}}{|q_{f}B|}e^{-\frac{u^{\prime}\epsilon}{|q_{f}B|}}e^{-\frac{iau^{\prime}}{|q_{f}B}|}\left(\cot(u^{\prime})-\frac{1}{u^{\prime}}\right)\,.
\label{I_Integral_Closed_Prime}
\end{equation}
It is easy to see that the integral  over $C_{2}$ vanishes when $R\to\infty$, due to the exponential damping in Eq.~(\ref{I_Integral_Closed_Prime})
\begin{equation}
I_{C_{2}}=\int_{C_{2}}\frac{du^{\prime}}{|q_{f}B|}e^{-\frac{u^{\prime}\epsilon}{|q_{f}B|}}e^{-\frac{iau^{\prime}}{|q_{f}B|}}\left(\cot(u^{\prime})-\frac{1}{u^{\prime}}\right)=0\,.
\label{I_C2}
\end{equation}
Now, for $C_{3}$ it is convenient to make the following change of variable
\begin{equation}
u^{\prime}=-i\omega^{\prime}\,.
\end{equation}
Thus, the integral over $C_{3}$ becomes
\begin{equation}
\begin{split}
I_{C_{3}}&=\int_{C_{3}}\frac{du^{\prime}}{|q_{f}B|}\exp\left(-\frac{u^{\prime}\epsilon}{|q_{f}B|}\right)\exp\left(-\frac{iau^{\prime}}{|q_{f}B|}\right)\\&\times\left(\cot(u^{\prime})-\frac{1}{u^{\prime}}\right)\\&=-\int_{0}^{\infty}\frac{d\omega^{\prime}}{|q_{f}B|}\exp\left(\frac{i\epsilon\omega^{\prime}}{|q_{f}B|}\right)\exp\left(-\frac{a\omega^{\prime}}{|q_{f}B|}\right)\\&\times\left(\coth(\omega^{\prime})-\frac{1}{\omega^{\prime}}\right)\,.
\end{split}
\end{equation}
Since $\epsilon\to0$, the imaginary exponential above tends to $1$, and the integral turns out to be analytic; the result is given by
\begin{equation}
I_{C_{3}}=\frac{1}{|q_{f}B|}\left(-\frac{|q_{f}B|}{a}+\ln\left(\frac{a}{2|q_{f}B|}\right)-\psi^{(0)}\left(\frac{a}{2|q_{f}B|}\right)\right)\,,
\label{I_C3}
\end{equation}
where $\psi^{(0)}(z)$ is the digamma function.\\
Substituting Eqs.~(\ref{I_C2}) and (\ref{I_C3}) in Eq.~(\ref{I_Integral_Closed_Prime}) and using Cauchy's residue theorem in the closed integral $I_{C}$, we can obtain the integral along the path $C_{1}$ with the result
\begin{widetext}
\begin{eqnarray}
\!\!\!\!\!\!\!\!\!\!\!\!{\mbox{PV}}\,(I)&=&{\mbox{PV}} \int_{0}^{\infty}\frac{du^{\prime}}{|q_{f}B|}e^{-\frac{u^{\prime}\epsilon}{|q_{f}B|}}e^{-\frac{iau^{\prime}}{|q_{f}B|}}\left(\cot(u^{\prime})-\frac{1}{u^{\prime}}\right)\nonumber\\
\!\!\!\!\!\!\!\!\!\!\!\!&=&\!\frac{i\pi}{|q_{f}B|}\!\Bigg(\frac{1}{-1\!+\!e^{(ia+\epsilon)\frac{\pi}{|q_{f}B|}}}\!+\!\frac{\ln\left[1\!-\!e^{-(ia+\epsilon)\frac{\pi}{|q_{f}B|}}\right]}{\pi}\Bigg)\!+\!\frac{1}{|q_{f}B|}\!\left(\!-\frac{|q_{f}B|}{a}\!+\!\ln\left(\!\frac{q}{2|q_{f}B|}\!\right)\!-\!\psi^{(0)}\left(\!\frac{a}{2|q_{f}B|}\!\right)\!\right).
\label{G1_Contribution}
\end{eqnarray}
Now we concentrate on the part of the integral in $F(0,p^{2}_{3},0,|q_{f}B|,m_{f})$ that comes from the term $-iG_{2}$, namely
\begin{equation}
I_{G_{2}}\equiv \frac{ig^{2}}{\pi^{2}}|q_{f}B|\int_{0}^{1}dv\int_{0}^{\infty}due^{-u\epsilon}e^{-iau}\Bigg[|q_{f}B|\csc^{2}(|q_{f}B|u)+\frac{\cot(|q_{f}B|u)}{u}-\frac{2}{|q_{f}B|u^{2}}\Bigg]\,.
\label{G2_Integral}
\end{equation}
Let us isolate the $u$ integral defining
\begin{equation}
J=\int_{0}^{\infty}due^{-u\epsilon}e^{-iau}\Bigg[|q_{f}B|\csc^{2}(|q_{f}B|u)+\frac{\cot(|q_{f}B|u)}{u}-\frac{2}{|q_{f}B|u^{2}}\Bigg]\,.
\label{J_Integral}
\end{equation}
It is again convenient to make the change of variable $u^{\prime}=|q_{f}B|u$ so that the $J$ integral becomes
\begin{equation}
J=\int_{0}^{\infty}du^{\prime}e^{-\frac{u^{\prime}\epsilon}{|q_{f}B|}}e^{-\frac{iau^{\prime}}{|q_{f}B|}}\Bigg[\csc^{2}(u^{\prime})+\frac{\cot(u^{\prime})}{u^{\prime}}-\frac{2}{{u^{\prime}}^{2}}\Bigg]\equiv J_{1}+J_{2}\,,
\label{J_Integral_Prime}
\end{equation}
where we have defined
\begin{subequations}
 \begin{align}
 J_{1}  = &  \int_{0}^{\infty}du^{\prime}e^{-\frac{u^{\prime}\epsilon}{|q_{f}B|}}e^{-\frac{iau^{\prime}}{|q_{f}B|}}\Bigg[\csc^{2}(u^{\prime})-\frac{1}{{u^{\prime}}^{2}}\Bigg]\,,\\
  J_{2}  = & \int_{0}^{\infty}du^{\prime}e^{-\frac{u^{\prime}\epsilon}{|q_{f}B|}}e^{-\frac{iau^{\prime}}{|q_{f}B|}}\Bigg[\frac{\cot(u^{\prime})}{u^{\prime}}-\frac{1}{{u^{\prime}}^{2}}\Bigg] \,.
 \end{align}
 \label{J1_J2}
\end{subequations}
$J_{1}$ can be integrated by parts to bring it to a form similar to the $I$ integral in Eq.~(\ref{I_Integral}). Taking into account the $-i|q_fB|$ factor in Eq.~(\ref{G2_Integral}), we find that
\begin{equation}
(\!-i|q_fB|)\,
{\mbox{PV}} \,(J_{1})\!=\!-(ia+\epsilon)\Bigg\{\!\pi\Bigg[\frac{1}{\!-1\!+\!e^{(ia+\epsilon)\frac{\pi}{|q_{f}B|}}}+\frac{\ln\left[1\!-\!e^{-(ia+\epsilon)\frac{\pi}{|q_{f}B|}}\right]}{\pi}\Bigg]\!-\!i\!\left(-\frac{|q_{f}B|}{a}\!+\!\ln\!\left(\frac{q}{2|q_{f}B|}\!\right)\!-\!\psi^{(0)}\left(\!\frac{a}{2|q_{f}B|}\!\right)\!\right)\!\!\Bigg\}\,.
\label{J1_Principal_Value}
\end{equation}
Finally, $J_{2}$ is calculated in a similar fashion as was done to compute the $I$ integral, namely, by promoting it to a closed contour integral using the same contour of integration. The result for $J_{2}$, taking into account again the factor $-i|q_{f}B|$ from Eq.~(\ref{G2_Integral}), is given by
\begin{eqnarray}
(-i|q_fB|)\,{\mbox{PV}}\,(J_{2})&=&\pi |q_{f}B|\Bigg(-\frac{\ln\left[1-e^{-(ia+\epsilon)\frac{\pi}{|q_{f}B|}}\right]}{\pi}-\frac{Li_{2}\left(e^{-(ia+\epsilon)\frac{\pi}{|q_{f}B|}}\right)}{\pi^{2}}\Bigg)\nonumber\\
&+&|q_fB|\ln\left(\frac{a}{|q_fB|}\right)-a\ln\left(\frac{a}{2|q_{f}B|}\right)+a+2|q_fB|\ln\left(\Gamma\left[\frac{a}{2|q_{f}B|}\right]\right)-|q_{f}B|\ln(4\pi)\,.
\label{J2_Principal_Value}
\end{eqnarray}
Putting together the expressions for $I$, $J_{1}$, and $J_{2}$, we get
\begin{equation}
\begin{split}
F(0,q^{2}_{3},0,|q_{f}B|,m_{f})=&-\frac{4g^{2}}{(4\pi)^{2}}\int_{0}^{1}dv\\&\times\Bigg\{-2i\pi v(1-v)q^{2}_{3}\Bigg[\frac{1}{-1+e^{(ia+\epsilon)\frac{\pi}{|q_{f}B|}}}+\frac{\ln\left[1-e^{-(ia+\epsilon)\frac{\pi}{|q_{f}B|}}\right]}{\pi}\Bigg]\\&-2v(1-v)q^{2}_{3}\left(-\frac{|q_{f}B|}{a}+\ln\left(\frac{a}{2|q_{f}B|}\right)-\psi^{(0)}\left(\frac{a}{2|q_{f}B|}\right)\right)\\&+\pi |q_{f}B|\Bigg(-\frac{\ln\left[1-e^{-(ia+\epsilon)\frac{\pi}{|q_{f}B|}}\right]}{\pi}-\frac{Li_{2}\left(e^{-(ia+\epsilon)\frac{\pi}{|q_{f}B|}}\right)}{\pi^{2}}\Bigg)\\&+|q_{f}B|\ln\left(\frac{a}{|q_{f}B|}\right)-a\ln\left(\frac{a}{2|q_{f}B|}\right)+a+2|q_{f}B|\ln\left(\Gamma\left[\frac{a}{2|q_{f}B|}\right]\right)-|q_{f}B|\ln(4\pi)\Bigg\}\,.
\label{Final_Expression}
\end{split}
\end{equation}
Finally, taking the real part of the previous expression, we obtain
\begin{eqnarray}
\Re [F(0,q^{2}_{3},0,|q_{f}B|,m_{f})]&\!\!=\!\!&-\frac{4g^{2}}{(4\pi)^{2}}\int_{0}^{1}dv\Bigg\{-2v(1-v)q^{2}_{3}\Bigg[A_{1}+A_{2}\Bigg]\nonumber\\
&\!\!+\!\!&|q_{f}B|\Bigg[\frac{\epsilon\pi}{2|q_{f}B|}\!-\!\ln\Bigg(\sqrt{2\cosh\left(\frac{\epsilon\pi}{|q_{f}B|}\right)\!-\!2\cos\left(\frac{a\pi}{|q_{f}B|}\right)}\Bigg)\Bigg]\!-\!\frac{|q_{f}B|}{\pi}\Bigg[\Re\Bigg(Li_{2}\left(e^{-(ia+\epsilon)\frac{\pi}{|q_{f}B|}}\right)\Bigg)\Bigg]\nonumber\\
&\!\!+\!\!&|q_{f}B|\ln\left(\frac{a}{|q_{f}B|}\right)-a\ln\left(\frac{a}{2|q_{f}B|}\right)+a-|q_{f}B|\ln(4\pi)+2|q_{f}B|\ln\Gamma\left(\frac{a}{2|q_{f}B|}\right)\Bigg\}\,,
\label{Final_Expression_Re}
\end{eqnarray}
where $A_{1}$ and $A_{2}$ are given by
\begin{align}
     A_{1}&=\Bigg[\frac{\pi}{2}\frac{\sin(\frac{a\pi}{|q_{f}B|})}{\cosh(\frac{\epsilon\pi}{|q_{f}B|})-\cos(\frac{a\pi}{|q_{f}B|})}-\tan^{-1}\Bigg(\frac{e^{\frac{\epsilon\pi}{|q_{f}B|}}\sin(\frac{a\pi}{|q_{f}B|})}{1-e^{\frac{\epsilon\pi}{|q_{f}B}|}\cos(\frac{a\pi}{|q_{f}B|})}\Bigg)\Bigg],\nonumber\\
     A_{2}&=-\frac{|q_{f}B|}{a}+\ln\left(\frac{a}{2|q_{f}B|} \right)-\psi^{(0)}\left(\frac{a}{2|q_{f}B|}\right)\,. 
\label{A_Bfinal}
\end{align}
Eq.~(\ref{Final_Expression_Re}) corresponds to the result in section~\ref{sec4} given by Eq.~(\ref{previoustoAB}).
\end{widetext}
\end{appendix}


\end{document}